# INTRINSIC LOAD-RESISTING CAPACITY OF KINESIN

Wenwei Zheng[1], Dagong Fan[1], Min Feng[3], Zhisong Wang[2]*

[1]Institute of Modern Physics, Fudan University, Shanghai 200433, China

[2]Physics Department, and Center for Computational Science and Engineering, National University of Singapore, Singapore 117542

[3]Shanghai Cancer Institute, Shanghai 200032, China

*Corresponding author (Email: phywangz@nus.edu.sg)

Contact details:

Zhisong Wang

Associate Professor

Department of Physics, National University of Singapore

2 Science Drive 3, Singapore 117542

Tel: (65)-6516-2606, Fax: (65)-6777-6126 Email: phywangz@nus.edu.sg

Webpage: http://www.physics.nus.edu.sg/corporate/staff/wangzhisong.html

**Abstract**

Conventional kinesin is a homodimeric motor protein that is capable of walking unidirectionally along a cytoskeletal filament. While previous experiments indicated unyielding unidirectionality against an opposing load up to the so-called stall force, recent experiments also observed limited processive backwalking under superstall loads. This theoretical study seeks to elucidate the molecular mechanical basis for kinesin's steps over the full range of external loads that can possibly be applied to the dimer. We found that kinesin's load-resisting capacity is largely determined by a synergic ratchet-and-pawl mechanism inherent in the dimer. Load susceptibility of this inner molecular mechanical mechanism underlies kinesin's response to various levels of external loads. Computational implementation of the mechanism enabled us to rationalize major trends observed experimentally in kinesin's stalemate and consecutive back-steps. The study also predicts several distinct features of kinesin's load-affected motility, which are seemingly counterintuitive but readily verifiable by future experiment.

# 1. Introduction

Conventional kinesin is a homodimer motor protein that transports organelles along cytoskeletal filaments called microtubules (MTs). Conventional kinesin consists of two catalytic core domains (often termed "heads") containing nucleotide- and MT-binding sites[1]. The catalytic core and the coiled coil dimerization domain are connected by a short peptide of ~ 14 residues, termed a "neck linker"[2]. A kinesin dimer is able to walk unidirectionally towards the plus-end of MT by a fixed step size of 8nm per ATP consumption[3, 4]. Kinetic measurements[5-7] revealed that the two heads of a kinesin dimer alternately hydrolyze ATP. The early kinetic studies also implied a hand-over-hand walking gait, which was later confirmed by a single-molecule fluorescence spectroscopy measurement[8]. This kinetic mechanism of alternate catalysis is consistent with the observation that a kinesin dimer can make as many as hundreds of consecutive steps without falling off MT[9].

A remarkable load-resisting capacity of individual kinesin dimers has been found by single-molecule mechanical measurements[10-16] in which an opposing force is applied to the dimerization neck domain. A kinesin dimer is able to keep running towards the MT plus-end against an opposing force of several picoNewtons[10-15]. Only when the force is raised to a critical value (i.e. the so-called stall force) of 6-8 pN, is the dimer halted completely[10-12, 14, 15]. The dimer appears to stay in this stalemate state until it falls off entirely from MT. Early single-molecule mechanical measurements[11-14] did not report consecutive backward steps when the load was raised beyond the stall force. It came as a surprise that Carter and Cross[15] observed processive backward walking of kinesin dimers by suddenly raising the load to superstall loads of 12 – 15 pN. More recently, Vale and coworkers[16] observed processive backwalking at zero ATP for opposing loads as low as 3 pN. Compared to normal forward walking under low loads[9], the consecutive backward run length is very short, and the average velocity extremely low.

This apparent disparity between forward and backward walking, together with the high value for the stall force, reflects an extraordinary inherent capability of individual kinesin dimers to resist opposing loads.

Controversies exist concerning the load-resisting capacity of kinesin. Firstly, the mechanochemical details of the stalemate state remains unclear. Some early single-molecule measurements[11]-[13] and also the recent experiment of Carter and Cross[15], found that the stall force changes little as the ATP concentration is increased by almost one hundred fold. Carter and Cross suggested that the stalemate state is a single-head binding configuration (so-called tethered state), in which the MT-bound head is waiting for ATP binding. A stall force measurement by Visscher, Schnitzer and Block[14] using both the commonly used fixed trap setup as well as the feedback-controlled position clamp apparatus found that the stall force increases with ATP concentration. More specifically, the position clamp data showed an increase in the stall force from 5.5 pN to 8 pN as the ATP concentration was increased from 5 $\mu$M to 2 mM. Theoretical studies[17-21] for kinesin motility often predicted a velocity-load curve first, and then used the criterion of zero average velocity to determine the stall force. Some theories[17, 19, 21] predicted no ATP dependence of the stall force. Other studies[18, 20] predicted an ATP-dependent stall force, but it is unclear how well these theories can explain the extremely low backwalking velocity and processivity at superstall loads.

Secondly, the mechanism for kinesin's backwalking under superstall loads is unsolved. Carter and Cross observed an ATP dependence of the dwell times for backward steps[15]. They proposed a stepping mechanism that is different from the one[1] suggested by earlier studies. In the model of Carter and Cross, both forward and backward stepping requires ATP binding to the MT-bound head. However, they found no evidence that the backward stepping motion was coupled to ATP turnover.

The distinct disparity between the velocities of kinesin's forward and backward walking hints at the existence of some direction-locking mechanism[12] inherent in the kinesin dimer. A process termed neck linker zippering[22], in which the neck linker is docked to the catalytic core and pointed towards the MT plus-end upon ATP binding, has been suggested to be the putative locking mechanism[1]. The suggestion was called into question[23, 15] later because the neck linker zippering has a free-energy gain of merely ~ 1.2 $k_B T$ ( $k_B$ is the Boltzmann constant and $T$ is absolute temperature)[24]. Recently we proposed from the physical principles a molecular mechanical mechanism [25, 26] by which the local conformational change of liker zippering is amplified into a directional locking effect for kinesin dimer as a whole. Interestingly, the seemingly trivial amount of zippering free energy turns out to be sufficient for a robust directional locking. This physical mechanism, termed the molecular ratchet-and-pawl mechanism, also facilitates a fine coordination between the two identical motor domains in line with the kinetic model of alternate catalysis. This molecular ratchet-and-pawl mechanism is essentially a synergic mechanism, in which the length of the neck linker plays a critical role. It is not surprising that the molecular ratchet-and-pawl mechanism also offers concrete guidelines for the designs of kinesin-mimicking artificial nanomotors[27].

In this article, we shall extend the ratchet-and-pawl mechanism into the superstall regime, and seek a unified molecular mechanical mechanism for kinesin's load-resisting capacity over the full range of external loads that can be possibly applied to kinesin dimer. Extensive computational studies will be carried out, with a focus on the stalemate state and the processive backwalk under superstall loads.

**2. Synergic, molecular mechanical mechanism for kinesin's directionality**

**2.1 Molecular ratchet-and-pawl mechanism under low loads**

The ratchet-and-pawl mechanism, as identified in our previous study[25, 26], selects and locks a kinesin dimer's movement into the MT plus-end. This molecular-physical mechanism for kinesin's unidirectionality has been derived from a molecular mechanical calculation for the free energies of the neck linkers, which are in a stretched conformation when both heads of the dimer are bound to MT. The ratchet-and-pawl mechanism can be understood by examining the energetic hierarchy of the dimer-MT binding configurations shown in Fig. 1 (See Method section for computational details). When the opposing force is under ~ 18 pN, the lowest-energy dimer-MT configuration is the double-head binding configuration marked as configuration I, in which the rear head is accompanied by a zippered linker and the front head is zippering-free. When a dimer interacts with MT, the lowest-energy configuration I will occur most frequently according to Boltzmann's law. The lowest-energy configuration is asymmetric in the sense that neck linker zippering occurs at the rear head but not at the front head. Therefore, ATP hydrolysis readily occurs at the rear head and initiates the head detachment. During the diffusion of the detached head, the former front head is allowed to bind ATP. The ensuing linker zippering at the standing head will bias the mobile head's diffusion towards the MT plus-end. If the diffusing head binds successfully to a forward site, a normal forward step occurs and the dimer resumes the lowest-energy configuration. If the diffusing head otherwise rebinds to its original position, which may occur if the standing head remains in a nucleotide-free state, a futile step occurs. In this case, the dimer also resumes the lowest-energy configuration. Again, it is the rear head rather than the front head that will ready itself to bind ATP and attempt to make another forward step. In summary, the kinesin dimer always moves towards the MT plus-end as if the movement were locked into this direction. Thus, the plus-end directionality of kinesin dimer arises from a synergic mechanism that involves a unique energetic hierarchy for the binding configurations of the entire dimer with MT. When the dimer is truncated into individual monomers, the synergy is destroyed. Monomeric kinesin, e.g. KIF1A still has limited processivity[28], but the unyielding plus-end directionality typical of the kinesin dimer is lost.

In addition, the vast energy gap in the configurational hierarchy imposes a coordination between the catalytic processes of the two heads. ATP binding and the ensuing zippering at the front head in the lowest-energy configuration I and the ATP-free two-headed binding configuration IV amounts to a transition to the high-lying configuration V and VI, respectively. Yet both of the energy gaps are much beyond the amount affordable by ATP hydrolysis[25]. As a consequence, zippering at the front head is virtually forbidden, if a tight coupling between ATP binding and zippering is assumed. Therefore, the hydrolysis-initiated detachment normally occurs for the rear head rather than the front head as long as the direction-locking mechanism is valid.

Thus, kinesin is essentially a molecular ratchet-and-pawl device, in which the asymmetric lowest-energy configuration serves the role of "ratchet" and the selective detachment of the rear head the role of "pawl". This ratchet-and-pawl mechanism has enabled us to rationalize the observed properties of kinesin's forward steps under pre-stall loads[25]. E.g. kinesin's configurational hierarchy and the gap-imposed coordination point straightforwardly to a major mechanochemical cycle for the dimer's steps under low load, which is illustrated in Fig. 2A. Suppose the dimer is initially in the most stable dimer-MT binding configuration I, in which the rear head is in an ATP-bound state with a zippered neck linker. ATP hydrolysis detaches the rear head, and cause a transition to configuration II. Prior to ATP binding to the MT-bound head, the diffusing head may bind to MT at a forward or backward site with no directional bias. However, the head-MT binding is relatively rare before ATP binding because of a non trivial energy barrier (~ 15 $k_B T$ [25]) caused by the self-stretch of the neck linkers. Consequently, when ATP binding becomes rate-limiting at low ATP concentrations, kinesin mainly stays in the single-headed binding configuration II. This agrees satisfactorily with a recent fluorescence experiment[29]. The event of ATP binding to the standing head then leads to configuration III. The neck linker zippering at the track-bound head biases the mobile head's diffusion towards the plus-end of the microtubule, because the zippering lowers down the forward-binding barrier by ~ 6 $k_B T$ [25] but raises the back-binding barrier drastically. Consequently, the zippering essentially locks the dimer into a forward step when the dimer is free of load. The binding of the diffusing head at a forward site causes the transition back to initial configuration I. This movement cycle is basically the same one proposed before by experimental biologists Vale and Milligan[1]. The above cycle for dimer motion dictates a hand-over-hand walking gait, which is an inherent character of the ratchet-and-pawl mechanism.

As can be seen in Fig. 1, kinesin's configurational hierarchy can be distorted by loads. We note that the normally lowest-energy configuration I (and other double-head binding configurations in general) is affected little by the load. This is because the position of the load-bearing coiled domain is shifted only by a small value in the stable double-head binding configurations. But energies for single-head binding configurations are drastically reduced by increasing loads. Notably, the energy for the single-head configuration with a zippering-free MT-bound head (marked as configuration II) decreases to cross the energy of a double-head configuration with both heads free of zippering (marked IV) as the load approaches ~ 8 - 9 pN. This critical load coincides with the upper limit of the stall forces found in the experiment of Visscher, Schnitzer and Block[14]. A load below this critical value is unlikely to result in the detachment of the load-bearing front head from the lowest-energy configuration I, because the energy gap between the double-head configuration I and the final single-head configuration is very large, ~ 10 $k_B T$ (see Fig. 1). Thus the load range of from zero to 8 - 9 pN is the regime where the direction-locking mechanism governs kinesin's forward walking.

## 2.2 Defect of the ratchet-and-pawl under high loads and onset of new pathways for head detachment

When the load is beyond the critical value of ~ 8 - 9 pN, the double-head binding configuration IV becomes even higher in energy than the single-head configuration II. Then, the unstable configuration IV readily decays to single-head binding by spontaneous detachment of the load-bearing front head. This pathway for front head detachment is purely mechanical in nature. Any mechanism leading to the detachment of the front head will render the direction-locking defective. This abolishes the "pawl" in the ratchet-and-pawl mechanism. Therefore, the load range beyond ~ 8 - 9 pN may be regarded as the regime of defective direction-locking.

In addition to the pure mechanical pathway described above, a very high load can also activate other pathways for front head detachment from double-head binding configurations. When the opposing force is large enough, the zippered neck linker at a head directly bearing a load can likely be untied from the catalytic core because of the low free-energy gain of the linker zippering. For example, such a mechanical unzippering probably occurs in a single-head binding configuration if the standing head is ATP-bound. Biased by the opposing force, the subsequent binding of the diffusing head at a rear site then leads to a new type of double-head binding configuration in which the front head is ATP-bound but zippering-free. The front head in such a state can detach either spontaneously under the load, or as a consequence of a delayed ATP hydrolysis. These two pathways for front head detachment are termed ATP-accompanied mechanical detachment and hydrolysis-initiated detachment, respectively.

In summary, when the load is raised into the superstall regime, three new pathways for front head detachment start to occur: the pure mechanical pathway, the ATP-accompanied mechanical pathway and the hydrolysis-initiated pathway. An illustration of these pathways is given in Fig. 2B.

**2.3 A unified molecular mechanics basis for kinesin's forward and backward walking**

The ratchet-and-pawl mechanism provides a unified ground for understanding kinesin's forward as well as backward walking. In this study we quantitatively modeled the response of kinesin's inherent ratchet-and-pawl mechanism to a broad range of loads. We performed a molecular mechanical calculation to determine the load-dependence of the ratchet-and-pawl mechanism, which in turn imposes restrictions on the transitions between dimer-MT configurations. Besides, the mechanical calculation yields barriers for head detachment and binding as a function of external loads. The mechanical calculation also yields the amount of force directly affecting individual MT-bound heads through adjacent neck linkers. The calculated forces were then used to determine the head's load-dependent catalytic rates. On the basis of the load-dependent ratchet-and-pawl mechanism, a kinetic Monte Carlo method was used to simulate kinesin's walking under pre−stall to superstall loads.

**3. Computational methods**

**3.1 Molecular mechanical calculations for dimer-MT binding configurations**

The total free energy of such a double-head binding configuration (E) is the sum of the neck linker free energy ($F_N$), the binding energies between kinesin heads and MT ($U_B$) and the free-energy gain by linker zippering ($U_Z$).

$$E = F_N + U_B + U_Z. \qquad (1)$$

The Helmholtz free energy of the neck linkers depends on the mechanical properties of the linker peptide as well as the geometry of the dimer-MT binding configuration. The liker free energy was calculated based on an interpolation formula for force-extension relationship of a worm-like chain. Validity of the worm-like chain formula has been repeatedly verified by mechanical measurements on single polymers. Major mechanical parameters of the neck linkers used in the worm-like chain formula (e.g. persistence length) were determined using a big ensemble of the linker peptide conformations that were generated by a Monte Carlo procedure. The geometrical parameters for the dimer-MT binding configurations, also required for the linker free-energy calculation, were taken from experimental data. Details for the linker free-energy calculation and for the parameter determination can be found in a previous publication and its appendix[25]. We used the experimentally determined values for the zippering energy and head-MT binding energies[30, 24]. Specifically, we used $U_Z$ = -1.2 $k_B T$ for the zippering energy. The head-MT binding energies are $U_B$ = -15 $k_B T$ for ADP-bound heads, and $U_B$ = -19 $k_B T$ for ATP-bound or nucleotide-free heads.

This study focuses on the cases in which a constant force is applied to the neck coiled coils. We determined the extensions of both linkers by balancing forces at the coiled domain. The Gibbs free energy for the neck linkers was calculated by combining the linkers' internal Helmholtz free energy with the contribution of the external force[31]. In the single-head binding configurations, the MT-bound standing head alone bears the opposing load, while in a double-head binding configuration, the neck linker adjacent to the front head is more extended than the linker adjacent to the rear head. Consequently the forces inflicted upon the two MT-bound heads by their adjacent neck linkers are different, and are given by derivatives of the Helmholtz free energies for the respective linker peptides. The calculated force for individual heads was used to consider the load-dependence of the kinesin's steps.

**3.2 Load-dependent rates**

In this study, we assumed an exponential form for the load-dependent rate for the front head detachment by the two mechanical pathways (i.e. the pure mechanical pathway and the ATP-accompanied pathway).

$$\gamma_d(F) = \gamma_d(F=0)\exp\{-\Delta E(F)/k_B T\}. \quad (2)$$

Here, $\Delta E$ is the free-energy barrier for the head detachment, which depends on the external load ($F$). The load-dependent barrier can be extracted from the dimer-MT configurational energies. For example, the barrier for the pure mechanical detachment is the energy difference between the double-head binding configuration IV and the single-head configuration II, i.e. $\Delta E = E_{IV}(F) - E_{II}(F)$ (see Fig. 1). The barrier for the ATP-accompanied mechanical detachment can be calculated in a similar way. A recent experiment by Guydosh and Block[32] suggested that ATP binding to a MT-bound head promotes the detachment of that head from MT. We assume that this effect reduces the detachment barrier calculated from the configurational energy difference by a certain amount $\Delta E'$. In this study, we determined the value of $\Delta E'$ by fitting the calculated dwell times to the experimental data. This procedure yields a value of $\Delta E'$ = 2.7 $k_B T$. From the experimental data of Guydosh and Block[32], an ATP-accompanied detachment rate of ~ 1.1 s$^{-1}$ can be deduced for the load $F$ = 3.9 pN. This result was used in determining the pre-exponential factor for the load-dependent rates of the pure mechanical detachment and ATP-accompanied mechanical detachment.

The hydrolysis-initiated detachment is a consequence of the combined processes of ATP hydrolysis and Pi release. The detachment rate depends on the loads, because Pi release may be affected by a rear-pointing force inflicted upon the head via the adjacent linker peptide. Specifically, we assumed that the hydrolysis-initiated detachment rate ($k_{hyd}$) and the ATP dissociation rate ($k_{off}$) depend on the rear-pointing force ($F_{head}$) by a Boltzmann-type relationship:

$$k_{hyd}(F_{head}) = k_{hyd}(F_{head}=0)/[p_1 + q_1 \exp(F_{head}\delta/k_B T)], \quad (3)$$

$$k_{off}(F_{head}) = k_{off}(F_{head}=0)\exp(F_{head}\delta/k_B T)/[p_2 + q_2 \exp(F_{head}\delta/k_B T)]. \quad (4)$$

Here $p_1 + q_1 = p_2 + q_2 = 1$. The above two equations were first introduced by Schnitzer, Visscher and Block in Ref.[9]. They conducted a global fit to experimental velocity-load data and found $q_1 = 0.0062$ ($p_1 = p_2$), which was used in this study. As for the value of $\delta$, we notice that Block and coworkers first reported a value of $\delta = 3.7$ nm[9], and then a revised value of 2.7 nm in a later article[33]. In this study, we found that the value of $\delta = 3.7$ nm leads to a better agreement on dwell times between our theory and the experiment. Thus, we used $\delta$ =3.7 nm throughout this study. We also found that the alternative value of 2.7 nm does not change the predicted values of the stall forces much.

We note that $F_{head}$ entering the above equations is the amount of force calculated for individual MT-bound heads (see the preceding subsection). For a single-head dimer-MT binding configuration, $F_{head}$ for the MT-bound head is equal to the external load ($F$) that is applied to the coiled coil dimerization domain. For a double-head kinesin-MT binding configuration, $F_{head}$ for either head does not equal the external load. This is a realistic treatment for the load effects on the kinesin's steps.

It is unlikely that the load applied to the coiled coil domain is sufficient to break the zippered conformation of the rear head when both heads are bound to MT, because the load is primarily borne by the front head. However, when a single head is bound to MT, unzippering at the MT-bound head will probably occur because of the low free-energy gain for the zippered conformation. In principle, when the load is increased, the population of the zippered conformation will decrease and that of the unzippered conformation will increase. In this study, we defined a threshold load above which the neck linker at the MT-bound head will remain permanently in the unzippered conformation. Without zippering at the MT-bound head, the bias for the mobile head's forward binding will be lost. Therefore, under a load beyond the threshold, forward stepping will hardly occur because of the limited length of the neck linkers. Thus, the value of the threshold load is close to the stall forces. In this study, we assumed a threshold load of 7.5pN. When the external load is above this threshold value and a single head is bound to MT, zippering at the MT-bound head is regarded to be completely forbidden.

**3.3 Kinetic Monte Carlo simulation**

We conducted a kinetic Monte Carlo simulation for kinesin's walking dynamics. For details of the simulation method, the reader is referred to Ref. [25]. In the simulation, we assumed that the release of $\gamma$-phosphate from the catalytic core triggers the detachment of the ADP-bearing head from the MT instantaneously. For double-binding configurations, the detachment of the ADP-associated head is likely assisted by the mechanical strain of the neck linkers[1]. ADP release from a head is assumed to occur upon its binding to MT. If the energy difference between two dimer-MT binding configurations is higher than the energy released from ATP hydrolysis, the transition between the two configurations is forbidden.

The rate for a diffusing head to bind MT was calculated using the first passage time theory[34-36]. The rate calculation considered the geometrical and energetic differences between the initial single-head binding configuration and the final double-head configuration[25]. The experimental values for the enzymatic rates of catalytic cores were used in this study. The following rates were taken from ref. [37] and references therein: ATP binding rate 3 $\mu M^{-1}$ $s^{-1}$, reverse dissociation rate 150 $s^{-1}$, hydrolysis rate 200 $s^{-1}$, rate for reverse ATP synthesis 25 $s^{-1}$, rate for $\gamma$-phosphate release 250 $s^{-1}$. The diffusion coefficient for head diffusion was taken as $3.5 \times 10^6$ $nm^2/s$.

**4. Results**

**4.1 Stall forces**

The simulation generated a large number of dimer trajectories for various ATP concentrations and opposing forces. The position of kinesin's dimerization neck domain as a function of running time was calculated from the trajectories of the two individual heads. By averaging the ensemble of trajectories, we obtained the average velocity as a function of opposing force. The stall forces were then determined as those opposing forces causing a zero average velocity.

In Fig. 3, we present the stall forces calculated from our simulation for various ATP concentrations. The calculations were conducted for constant forces, mimicking the measurement by Visscher, Schnitzer and Block using a feedback-controlled position clamp apparatus[14]. Thus, the calculated stall forces were compared with their position clamp data. The predicted and measured stall forces both increase with increasing ATP concentration. The present prediction and experimental data agrees well for low ATP concentrations, but the quantitative agreement is less satisfactory for saturating concentrations. This can probably be attributed to some force dependence, which becomes important at high ATP concentrations but is not taken into account by the present model.

**4.2 Dwell time and velocity**

Carter and Cross[15] measured the dwell times and average velocity for forward and backward steps separately using a new detection method. When they suddenly move MT away from the trap center, the force exerted on the dimer changes. When the dimer starts to make a step in response to such a load change, a step finder detects the direction for such a single step, and records its dwell time. The MT was moved once for each measurement[15]. To compare with the experimental data, we carried out a single step simulation in which a single step, either forward or backward, is recorded in a way mimicking the step-finder. In Fig. 4, we compare the predicted dwell times and velocity with the experimental data of Carter and Cross. As can be seen in the figure, the error bar is high for forward steps at high loads and for backward steps at low loads. This is because of the low stepping probability under these extreme conditions. We therefore present only the forward stepping results for loads below 10 pN, and the backward stepping results for loads above 6 pN.

The experimental data of Carter and Cross show that when the opposing force is raised, the dwell time first increases, and then slightly declines after reaching a peak value. However, the non-monotonic feature is less obvious for forward steps than for backward steps. The trend of dwell times is well reproduced by the present theory, though quantitative discrepancies exist between the experimental data and the theoretical predictions. The experiment found that the dwell time at a low ATP concentration of 10 $\mu$M is higher than that for a saturating concentration of 1mM. This ATP dependence is also reproduced by our theory. Notably, the dwell times for the forward and backward steps are not very different in magnitude for the same ATP concentration and external load. The present model has reproduced this experimental observation.

The data of Carter and Cross for a high ATP concentration (1 mM) show that the velocity drops more than half by increasing the opposing load from 0 to 1.5 pN (Fig. 4C). Previous experiments, notably those by Block et al.[9, 33], however found a far weaker load dependence at the low load range for saturating ATP concentrations. Carter and Cross[15] also found a dramatic increase of velocity when the load is directed towards the MT's plus end and increases from 0 to 5 pN. Again, the experiments by Block et al.[33], and more recently by Vale et al.[16] found however a slight increase of velocity with increasing assisting load. The prediction of the present model for the case of assisting loads (not shown here, see a previous study[26]) agrees with the experimental findings of Block et al. and Vale et al.. Apparently, it is impossible to quantitatively reproduce these contradictory data using a single set of molecular mechanical and rate parameters. This is hardly a surprise given the fact that Carter and Cross used Drosophila kinesin molecules for their single-molecule motility assay, and Block et al. and Vale et al. used squid or human kinesin. We therefore emphasize on a qualitative comparison instead of a quantitative one between theory and experiment.

In Fig. 4C, we present the velocity data over the load range of 0 – 15 pN from the Carter-Cross experiment together with predictions of the present model for two values of ATP concentration (10 $\mu$M and 1 mM). We also show in the figure the velocity data obtained by Block et al.[9]. Because the latter experiment did not report the force-velocity curve for the ATP concentration of 1 mM, we show the experimental curve for a higher concentration (2 mM). As Block et al. found in their experiment[9], the velocity data change only slightly over the range of saturating ATP concentrations. The experimental velocity-concentration curves[9] respectively obtained under an opposing load of 1.05 pN, 3.6 pN and 5.63 pN indicated that the velocity changes less than 15% for all of the three load values as ATP concentration increases from 1 mM to 2 mM. Therefore, the experimental data for 2 mM can be approximated to be those for 1 mM within an error of smaller than 15%, and compare meaningfully with the theoretical result for 1 mM.

For the low ATP concentration (1 $\mu$M), the present model agrees with the Carter-Cross data. For the high concentration (1 mM), the present model compares poorly with the Carter-Cross data except for the velocity at zero load, but agrees with the data of Block et al.. Overall, the present model qualitatively reproduces the shoulder-arm shape of the experimental velocity-force curves as the load increases from zero to the stall forces, and also reproduces the concave shape extending into the superstall regime. Both the experimental data and the theoretical results show extremely low values of back-walking velocity under super-stall loads. More importantly, the theory predicts a higher back-stepping velocity for 1 mM than 10 μM, in qualitative agreement with the Carter-Cross experiment. This ATP dependence is an evidence that the backward steps are coupled, at least partially with the hydrolysis cycle.

**4.3 Stalemate**

In the experiment of Visscher, Schnitzer and Block, only stall events that last no less than two seconds were qualified to be scored[14]. The trap stiffness of their apparatus is 0.037 pN/nm, and the error for force determination is ~ ± 0.1 pN. Then, the position uncertainty is ~ ± 3 nm (i.e. 0.1 pN/0.037 pN/nm ~3 nm). The average velocity at stall is less than ± 1.5 nm/s, which is negligible compared to the forward walking velocity at low loads. Evidently, Visscher, Schnitzer and Block[14] observed a genuine stalemate that sustains itself over a reasonable period of time (~ 2seconds). An early measurement by Coppin et al.[12] also found clues to such an stalemate state.

Can the present model predict the stalemate state? We note that previous theoretical studies[17-21] often rely on the sole criterion of zero average velocity to determine stall forces yet without specifying the details of the stalemate state. In principle, a zero average velocity can be caused by a state other than a real stalemate. For example, if the dimer keeps moving back and forth and forward steps cancel the backward ones, a zero net velocity will occur.

We examined the details of consecutive, effective stepping around stall forces as predicted by the present model. By "effective step" we mean a non-zero shift of the dimer's center-of-mass both starting from and ending at a double-headed dimer-MT binding configuration. In Fig. 5 B, we show predictions of the present model of the total number of consecutive, effective steps, both forward and backward. As can be seen clearly, the so-defined number of consecutive steps as a function of the opposing load exhibits a minimum at the respective stall forces for the values of ATP concentration shown (10 μM and 1 mM). The minimum dips to virtually zero effective step for both the low and high ATP concentrations, suggesting that the dimer under the stall forces barely makes any step either towards the plus end or minus end of the MT. Moreover, we found in the kinetic Monte Carlo simulation that the kinesin dimer under the stall force stays bound to MT for an average time well above one second. Therefore, the present model predicts a real stalemate of the dimer at the stall force.

**4.4 Walking behavior and backward run length**

In Fig. 5A, we present a typical trajectory for processive backwalking under an opposing force of 15 pN at an ATP concentration of 10 μM. This trajectory shows that the dimer makes several consecutive backward steps over a few seconds. These features are similar to the trajectories found by Carter and Cross (Fig. 1 of ref. [15]). Besides, the trajectory of backward walking shows a hand-over-hand gait similar to forward walking[8]. In the simulation, we found that any processive backwalking consisting of more than five consecutive backsteps is rare. This difficulty in observing consecutive backward steps is consistent with previous experiments[11, 12, 14, 15], and implies an extremely low processivity of backward stepping.

A close examination of the stepping details predicted by the present model found that at a given ATP concentration, the effective steps are predominantly towards the MT plus end (i.e. forward steps) for opposing loads below the stall force, and predominantly towards the minus end (i.e. backward steps) for opposing loads above the stall force. Therefore, the superstall part of the step number-load curves shown in Fig. 5B presents virtually the consecutive backward run length. The present model predicts two notable features of the backward stepping in the superstall regime. Firstly, the consecutive backward run length increases with increasing the superstall load, as if on the rebound from the minimum at the stall force. Secondly, at a given value of superstall load, the backward run length is improved by decreasing the ATP concentration. The second feature is largely a consequence of the first feature. For a lower ATP concentration, the stall force is smaller, and the rebound of the backward run length occurs at lower loads. Take a low concentration of 10 μM as an example, the backward run length starts to rise when the load is beyond the stall force of ~ 6.2 pN (see Fig.5B). When the load reaches ~ 9 pN, which is the theoretical stall force for a high concentration of 1 mM, the dimer will make virtually no effective step at this concentration. However, if the ATP concentration is lowered to 10 μM, consecutive backward steps will become possible at the same load (~ 9 pN). Overall, the ATP dependence of the superstall backwalk run length is weak, with an increase of less than 10% by increasing ATP concentration from zero to 1 mM.

As for the run length for forward steps below the stall force, the present model reproduces the experimentally observed trend[9] of decreasing run length with increasing loads (see the pre-stall part of the step number-load curves in Fig. 5B, and also a previous publication[26]). As compared to the pre-stall forward walking, the run length for supertsall backward walking predicted here is extremely poor. Overall, the average number of backward consecutive steps is below 2.5 for superstall loads of 8 – 15 pN (see Fig. 5B). The extremely poor processivity for backwalking under superstall loads is also hinted in experiment[12, 15, 16].

**4.5 Pathway analysis**

In Fig. 6, we compare occurrence probabilities of the three pathways for front head detachment during the dimer's walking under superstall loads. These simulation results show that the pure mechanical pathway and the ATP-accompanied mechanical pathway occur with an increasing likelihood as the load becomes higher. On the contrary, increasing loads reduce the occurrence of the hydrolysis initiated detachment via load-suppressed ATP turnover. As can be expected, both the ATP-accompanied and hydrolysis initiated pathways show a strong dependence on ATP concentration. At a high ATP concentration of 1 mM, the three pathways compete in the front head detachment. At an ATP concentration 100 fold lower (10 μM), the pure mechanical detachment dominates. Overall, more pathways contribute to backward walking as the ATP concentration is raised. This accounts for the lower dwell times for backward steps at higher ATP concentrations (Fig. 4B).

## 5. Discussions

### 5.1 Characteristics of stalemate

The present study predicts a distinct characteristic of kinesin's stalemate. Namely, a minimum in the number of consecutive effective steps, regardless of direction, occurs at the stall force when the opposing load is changed from low, pre-stall values to high superstall values. This characteristic leads to a prediction that the backwalking run length increases, though extremely slowly, by raising the superstall load. This load dependence is in sharp contrast to the forward walking, in which an opposing force generally reduces the run length as found experimentally[9], and predicted also by the present model (see Fig. 5B and also [26]). The seemingly counterintuitive load dependence of backward run length is however understandable, because a larger opposing load will more effectively direct the diffusive back-binding of a mobile head to MT, and reduces the chance of the MT-bound head's detachment from MT ( i.e. the entire dimer's derailment).

Because of the ATP dependence of the stall force, the above characteristic of a minimum at the stall force leads to a second prediction that at a given value of superstall load the backwalk run length is improved by decreasing ATP concentration (see RESULTS section). Again, this appears to be counterintuitive, and contrasts with the ATP dependence of the forward run length (i.e. higher forward run length for higher ATP concentrations), which was observed experimentally and correctly predicted by the present model (see the pre-stall part of the step number-load curves in Fig. 5B, and see also a previous publication[26]). The ATP dependence of backwalk run length is understandable too. Because kinesin's ratchet-and-pawl mechanism, particularly the direction-locking mechanism is rendered defect by the superstall loads, the fine coordination[5-7] between the two heads' hydrolysis cycles is largely lost. This explains the overall low backwalk run length predicted here and implicated in experiments[12, 15, 16]. The remnant of head-head coordination sustaining processive backwalking is mainly enforced by the external mechanical load. This mechanically enforced coordination will likely be interrupted by now largely uncorrelated nucleotide activities of individual heads, and will thus manifest itself better at low ATP concentration.

In consistency with the distinct ATP dependence, the present model predicts the highest backwalk run length at zero ATP concentration (see Fig. 5B). More interestingly, because the stall force at zero ATP is 0 pN, the present model predicts that processive back steps can occur at zero ATP even for low opposing loads. In a recent experiment, Vale and coworkers[16] indeed observed processive back steps for opposing loads as small as 3 pN. They reported a mean backwalk velocity of ~ 7.5 ± 1 nm/s for an opposing load of 3 ± 1.4 pN, and ~ 15 ± 2 nm/s for 6 ± 1.4 pN. The present model predicts 8.23 nm/s and 13.74 nm/s for the two values of loads, respectively, in fair agreement with the experimental data.

**5.2 ATP-dependence of stall forces**

The present model predicts an ATP-dependence of the stall force, in line with the experimental data of Visscher, Schnitzer and Block[14], but not with the data of Carter and Cross[15]. A recent theory by Shao and Gao[21] predicted no ATP dependence for either stall forces or dwell times for backward steps. Some other theories[18, 20] predicted ATP-dependent stall forces. However, these previous theories[21, 18, 20], particularly that of Kanada and Sasaki[20], tend to overestimate the processivity for backwalking. The present model predicts both an ATP-dependent stall force and an extremely low backwalking run length.

Based on the results of the present study, we may develop an understanding of the nucleotide dependence of the stalemate state as follows. In the stalemate state the probabilities for forward and backward stepping are balanced. The probability for a forward or backward step is mainly determined by two factors. The first factor is the head detachment, which occurs to the rear head for forward stepping, and to the front head for backward stepping. The second factor is the diffusive searching and binding of the detached head. The subsequent binding at a forward (backward) site results in a forward (backward) step. An analysis of ATP dependence of the two factors for forward and backward stepping will shed light on the nucleotide dependence of the stall forces.

The search-and-bind time of the diffusing head is affected by the nucleotide state of the other MT-bound head. When the standing head is ATP-bound and accompanied by a zippered neck linker, the mobile head is biased for forward binding. Consequently, the search-and-bind time becomes shorter for forward stepping and longer for backward stepping. So a higher ATP concentration favors forward walking as far as the search-and bind time is concerned. However, if the zippering is hindered under high loads, the search-and-bind time will become ATP-independent.

In a double-head dimer-MT binding configuration, the front head bears most of the opposing load, but the rear head is not directly affected by the load. The rear head detachment occurs predominately via

ATP hydrolysis, and thus is ATP-dependent for any loads. The front head detachment occurs by more than one pathway. The pure mechanical pathway is completely ATP-independent. The ATP-accompanied mechanical detachment and hydrolysis initiated detachment are ATP-dependent, but both detachment pathways can occur only when the load becomes sufficiently high. Under stalling loads, the rates for ATP binding and hydrolysis at the load-bearing front head are both drastically suppressed. Consequently, the ATP–dependent detachment of the front head is much slower than that of the rear head. Moreover, a detached rear head can rebind to the backward site again through the load-biased diffusion. This is because the load will exponentially reduce the search-and-bind time for a backward step, and exponentially prolong that for a forward step. Therefore, more than one ATP molecule can be consumed by the rear head if it undergoes repeated detachment and back-binding. Consequently, the effect of increasing ATP concentration on the rear head detachment is larger than on the front head detachment. Again, a higher ATP concentration favors forward walking.

Combining the ATP-dependence of the two factors, it becomes clear that an increase of ATP concentration increases the probability for a forward step more than the probability for a backward step. Consequently, a larger force is required to balance the forward and backward stepping to bring the dimer to a halt.

**5.3 Kinesin dimer is trapped in a self-closed loop formed by two fast processes under the stall force**

The mechanochemical processes underlying kinesin's stalemate can be understood by considering the rate-limiting steps for forward and backward walking under high loads. A forward step is a combination of two processes, i.e. rear head detachment and forward diffusive search-and-bind of the detached head.

This is schematic illustrated in Fig. 2A. Similarly, a backward step is a combination of front head detachment and backward diffusive search-and-bind, as illustrated in Fig. 2B. For a forward step under stalling loads, the diffusive search-and-bind is rate-limiting because the rate for this process is reduced exponentially by load. The hydrolysis-initiated rear head detachment initiating a forward step is almost load-independent, and therefore is a relatively fast process. For a backward step under stalling loads, the front head detachment is rate-limiting. Comparatively, the subsequent search-and-bind process to complete the backward step is a fast process, because it is exponentially accelerated by the load. The dimer's state under the stalling forces is dominated by the two fast processes, namely the hydrolysis-initiated rear head detachment and the backward diffusive binding. The combination of these two fast processes can form a repetitive loop in which the rear head detaches and then rebinds to the backward site again. Under the stalling loads, the dimer can be trapped in such a closed loop because the mobile head's back-binding to MT followed by ATP binding to the head will lead to the lowest-energy dimer-MT configuration (see Fig. 1). As a consequence, ATP can still be consumed but the entire dimer does not change its position. The closed loop provides the physical basis for a sustained stalemate, whose lifetime depends on the competition between the fast and slow processes. We note that the molecular events of head detachment and binding are stochastic processes. The above discussions of these processes being fast or slow must be understood in an average sense.

We note that the stalemate state proposed here is different from that suggested by Carter and Cross[15]. In their model, the stalemate state is primarily a single-head tethered state, while the stalemate state suggested here also contains a non trivial component of two-head binding configurations. A recent experiment[29] indicates that double-head dimer-MT binding configurations indeed dominate at high ATP concentrations. Furthermore, in the model of Carter and Cross, the rate-limiting process for backward stepping is ATP-binding to the MT-bound head, which ends the tethered state. In the present model, the backward stepping is instead rate-limited by detachment of the front head from MT.

**5.4 Load- and ATP-dependence of the dwell times**

Both this theory and the experimental data found that the dwell times for forward and backward steps both reach a maximum at stall forces. This feature can be understood again by considering the rate-limiting processes for forward and backward steps under the stalling forces. The stalemate state, namely the dimer-trapping loop formed by the two fast processes can be broken by occurrence of either of the two slow, rate-limiting processes. For example, once the rate-limiting process of forward search-and-bind occurs, a forward step is accomplished. Similarly, a backward step will be initiated if the rate-limiting process of front head detachment occurs. Because the two processes leading out of the dimer-trapping loop are both slow processes, the dwell time for either a forward or a backward step under stalling loads likely contains the time for repeated dimer-trapping loops. Consequently, the dwell times are generally high for stalling loads.

As found from an experimental study[14], the consumption of more than one ATP molecule per step is rare under pre-stall loads. This feature has been quantitatively reproduced by a previous study based on the ratchet-and-pawl mechanism[25]. The rareness of the dimer-trapping loop at low loads explains the low dwell times before stall. Under superstall loads, there also exists limitation to the time duration of the dimer-trapping loop. The entire dimer will fall off from the MT if detachment occurs in a single-head dimer-MT binding configuration. Under superstall loads, consecutive steps can occur only when the dwell times for these steps are short enough to avoid fall-off. Therefore, the fall-off rate imposes an upper limit to the dwell times of actually occurring consecutive steps. This explains the slow decrease of the dwell times with increasing superstall loads (Fig. 4).

The ATP dependence of the dwell times can be understood from our model. A higher ATP concentration obviously causes a shorter dwell time for forward steps, which are driven by ATP hydrolysis at the rear head. At a higher ATP concentration, a backward step can occur only if it takes a time short enough to compete with the fast forward stepping. Consequently, the average dwell time for successful backward steps decreases with increasing ATP concentration.

**5.5 Correlation between forward and backward dwell times**

Generally speaking, forward stepping and backward stepping always compete to occur at any ATP concentration and under any load. At low opposing loads, the probabilities for backward and forward stepping are undoubtedly disproportionate to each other. Nevertheless, a backward step must either be fast enough to compete with or simply be replaced by a forward step. In other words, the rare events of backward stepping actually observed in experiment at low loads are merely a subgroup of theoretically possible backward steps "filtered" by the dominant forward stepping events. Therefore, as far as the quickness of the stepping events is concerned, no large disparity is observable between forward and backward steps. This explains the experimental finding that the dwell times for forward and backward steps are similar in magnitude over a broad range of ATP concentrations and loads (Figs. 4A and B). The correlation between the dwell times for forward and backward steps have previously been suggested by Kolomeisky, Stukalin and Popov[38], albeit from a different perspective.

**5.6 Mechanochemical cycles for consecutive backwalking under superstall loads**

Continual backward steps require repeatable mechanochemical cycles. This study suggests that processive backwalking under superstall loads is likely caused not by a single self-closed cycle, but rather by multiple cycles between which the dimer can switch during the backward walking. Such multiple cycles for backward walking can be formed by the three pathways for front head detachment combined with the load-assisted back-binding. Following the illustrations in Fig. 2B, several backwalking cycles can be immediately identified. Specifically, a pure mechanical detachment from the double-head dimer-MT configuration IV followed by a back-binding can form a self-closed cycle for backward steps. Another cycle can be formed by a hydrolysis-initiated detachment followed by ATP binding to the MT-bound head. This cycle is similar to the cycle proposed by Carter and Cross[15]. Furthermore, an ATP-accompanied mechanical detachment followed by back-binding of the ATP-carrying mobile head will lead to the lowest-energy dimer-MT configuration I, which plays a major role in the movement cycle for forward walking.

The data of Carter and Cross [15] imply that backward steps are not related to ATP turnover, though ATP binding to the MT-bound head is evidently involved. In the present model, front head detachment triggers a backward step to occur by any one of three pathways. Both the hydrolysis-initiated detachment pathway and the ATP-accompanied mechanical detachment pathway depend on ATP binding (and thereby ATP concentration), but only the former pathway is related to ATP turnover. However, the occurring probability of the hydrolysis-initiated pathway drops to a negligible level when the load is raised to the superstall level (see Fig. 6B). Consequently, the processive backwalking under superstall loads is influenced by ATP concentration but not by turnover. Furthermore, none of the backwalking cycles identified above are a simple reversal of the forward cycle and lead to ATP re-synthesis. This conclusion is consistent with the experiment study of Hackney[39].

**5.7 Overall load-resisting capacity of kinesin**

An overall view of kinesin's intrinsic capacity to resist loads can be established by examining the distortion of the dimer-MT configurational hierarchy over a broad range of loads. The configurational energies presented in Fig. 1 expose three distinct regimes for kinesin's response to loads, although the boundaries between the regimes can be smeared by ubiquitous fluctuations. The first regime is for the load range from 0 pN to ~ 8 - 9 pN. As discussed already in section 2, this regime is where the ratchet-and-pawl mechanism works normally and the dimer's unidirectionality is well proved by experimental studies. The first regime may be termed the "work regime" for kinesin, in which the ratchet-and-pawl theory predicts the same movement cycle for the dimer's unidirectional walking as suggested previously by many authors (e.g. refs. [5-7, 1]). The second regime is for the load range from 8 - 9 pN to ~18 pN, in which Carter and Cross[15] observed consecutive backward steps. As discussed in preceding sections, the second regime is where the "pawl" mechanism (namely the selective detachment of the rear head from MT) is destroyed and direction-locking becomes defective. But the "ratchet" (namely the asymmetric lowest-energy dimer-MT configuration) is preserved. We term this regime the "buffer" regime for kinesin.

The third regime is for loads beyond ~ 18 pN. As can be clearly seen in Fig. 1, no double-head dimer-MT binding configuration is more stable than the single-head configuration II. From an energetic point of view, the transition from the single-head configuration to any double-head configuration cannot occur without energy input. Consequently, the binding of a mobile head to MT is virtually impossible even though the other MT-bound head is in an ATP-bound, zippered state. Then the ground for any meaningful step is lost altogether, because derailment of the entire dimer readily occurs from the dominant single-head configurations. The third regime for kinesin thus may be termed "wreck regime". Due to the existence of the wreck regime, neither the backwalking velocity (Fig. 4 C) nor the backward run length (Fig. 5 B) can increase unlimitedly. When the load is raised into the wreck regime, the dimer's processive walking will be impossible and any discussion on velocity and processivity will be meaningless. Thus, the load-velocity and load-processivity curve as shown in Figs. 4 and 5, respectively, cannot be extrapolated further to the wreck regime. In a previous study[21], Shao and Gao also predicted a sudden change in kinesin's walking behavior at ~ 18 pN. They argued that kinesin keeps walking under loads higher than ~18 pN, but the walking gait will change from the hand-over-hand manner to the inchworm manner. Here, we suggested that neither forward nor backward walking is possible when the load is larger than ~18 pN.

The existence of the middle regime (~ 8 – 18 pN) might have some biological implications. This regime might provide a protection mechanism against occasional mechanical hindrances, which presumably occur in the crowded interior of the cell. If the hindrance is temporary, the dimer may avoid total derailment by retreating a few tens of nanometers (corresponding to several consecutive backward steps) along MT. After the hindrance disappears, the dimer will resume its running towards the MT plus-end. If the hindrance lasts longer than a critical time, the dimer likely derails from MT. This critical time can be defined as the average number of consecutive backward steps multiplied by the average dwell time for such steps. Based on the experimental data for the dwell times of Carter and Cross[15] (Fig. 4 B) and on our prediction for the processivity (Fig. 5 B), we can estimate an upper limit for the critical time as 2 second for the load range of ~ 8 – 18 pN. Under short-lived mechanical hindrances (shorter than ~ 2 seconds), a dimer's transportation is delayed but not completely terminated. This is why we called the regime of ~ 8 – 18 pN the "buffer regime". If the mechanical hindrance is beyond the buffer regime (higher than ~ 18 pN), derailment readily occurs. Quick derailment in the wreck regime might be functionally meaningful too, because it prevents the dimer from sticking to the MT long enough to block the path for other motor proteins.

## 6. Conclusions

We have proposed a unified molecular mechanical mechanism for kinesin's response to the full range of external mechanical loads that may possibly applied in single-molecule experiments. Compared to previous theoretical studies, the present study is unique in that the molecular mechanical mechanism proposed facilitates directional rectification and fine head-head coordination from physical principles, and the load susceptibility of this inner mechanism determines kinesin's response to various levels of external loads. Numerical calculations within the framework of the mechanism have been carried out to gain quantitative understanding of kinesin's load-resisting capacity in general, with an emphasis on stalemate state under a stall force and limited processive steps under sperstall loads. Major findings can be summarized as below:

(1) The overall load-resisting capacity of a kinesin dimer is determined by a synergic ratchet-and-pawl mechanism. When the load is below ~ 8 - 9 pN, the mechanism works normally and locks the dimer's movement into the MT plus-end. When the load is raised to the superstall range from 8 – 9 pN to ~18 pN, the "pawl" mechanism is destroyed, but the "ratchet" mechanism remains intact. The defective ratchet-and-pawl mechanism allows a limited processive backwalking of the dimer. But the backwalk run length is extremely low, in line with the overall observations of recent experiments. When the load is further raised beyond ~ 18 pN, the dimer readily derails from MT, rendering impossible any processive walking.

(2) This study predicts a distinct feature that, if the total number of effective steps, both forward and backward, made consecutively by a kinesin dimer is measured as a function of opposing load, a minimum step number will appear at the stall force. Moreover, the minimum step number is virtually zero for a broad range of ATP concentrations. The present study thus concludes that the kinesin dimer under the stall force is trapped in a genuine stalemate that sustains itself for a period of time well above one second. The present study also predicts that the stall force depends on ATP concentration, in quantitative agreement with a previous experiment.

(3) As for the backwalking under superstall loads, this study predicts two features that contrast sharply the forward walking at low loads, and are seemingly counterintuitive. The first prediction is that the backwalk run length increases slightly by increasing the opposing load at a given ATP concentration. The second prediction is that at a given value of superstall loads, the backwalk run length improves by decreasing ATP concentration. Both predictions are consistent with the distinct feature of minimum step number at stalemate, and may be verified by future experiments. As a matter of fact, both predictions suggest that processive backward steps are most readily observed in the superstall regime (i.e. between the stall force and the upper limit of ~ 18 pN) and at low ATP concentration. This is in line with reported experimental studies on kinesin's processive backwalking. Interestingly, the predictions also suggest that consecutive backward steps can be observed at zero ATP for opposing load as low as 3 pN, in consistence with a recent experiment. The backwalking velocity at zero ATP predicted by this study agrees with the experimental data.

(4) This study confirms the experimental finding that kinesin's backwalking velocity under superstall loads is higher for higher ATP concentrations. An analysis suggests that consecutive backward steps are caused by multiple interconnected mechanochemical cycles. The experimentally observed trends of load- and ATP-dependencies of backwalk velocity and dwell times are rationalized based on the multiple cycles for backward steps.

**Acknowledgement**

We thank R. Cross for helpful discussions and for providing us the experimental data. This work was partly funded the staff startup fund of NUS.

**References**

[1]     Vale R D and Milligan R A 2000 The way things move: looking under the hood of molecular motor proteins *Science* 288 88-95

[2]     Kozielski F, Sack S, Marx A, Thormahlen M, Schoenbrunn E, Biou V, Thompson A, Mandelkow E-M and Mandelkow E 1997 The crystal structure of dimeric kinesin and implications for microtubule-dependent motility *Cell* 91 985-94

[3]     Hua W, Young E C, Fleming M L and Gelles J 1997 Coupling of kinesin steps to ATP hydrolysis *Nature* 388 390-3

[4]     Schnitzer M J and Block S M 1997 Kinesin hydrolyses one ATP per 8-nm step *Nature* 388 386-90

[5]     Hackney D D 1994 Evidence for alternating head catalysis by kinesin during microtubule-stimulated ATP hydrolysis *Proc. Natl. Acad. Sci. USA* 91 6865-9

[6]     Ma Y Z and Taylor E W 1997 Interacting head mechanism of microtubule-kinesin ATPase *J. Biol. Chem.* 272 724-30

[7]     Hancock W O and Howard J 1999 Kinesin's processivity results from mechanical and chemical coordination between the ATP hydrolysis cycles of the two motor domains *Proc. Natl. Acad. Sci. USA* 96 13147-52

[8]     Yildiz A, Tomishige M, Vale R D and Selvin P R 2004 Kinesin walks hand-over-hand *Science* 303 676-8

[9]     Schnitzer M J, Visscher K and Block S M 2000 Force production by single kinesin motors *Nat. Cell. Biol.* 2 718-23

[10]    Svoboda K and Block S M 1994 Force and Velocity Measured for Single Kinesin Molecules *Cell* 77 773-84

[11]    Meyhofer E and Howard J 1995 The force generated by a single kinesin molecule against an elastic load *Proc. Natl. Acad. Sci. USA* 92 574-8

[12]    Coppin C M, Pierce D W, Hsu L and Vale R D 1997 The load dependence of kinesin's mechanical cycle *Proc. Natl. Acad. Sci. USA* 94 8539-44

[13]    Kojima H, Muto E, Higuchi H and Yanagida T 1997 Mechanics of single kinesin molecules measured by optical trapping nanometry *Biophys. J.* 73 2012-22

[14]    Visscher K, Schnitzer M J and Block S M 1999 Single kinesin molecules studied with a molecular force clamp *Nature* 400 184-9

[15]    Carter N J and Cross R A 2005 Mechanics of the kinesin step *Nature* 435 308-12

[16]    Yildiz A, Tomishig M, Gennerich A and Vale R D 2008 Intramolecular strain coordinates kinesin stepping behavior along microtubules *Cell* 134 1030-41

[17]    Derenyi I and Vicsek T 1996 The kinesin walk: A dynamic model with elastically coupled heads *Proc. Natl. Acad. Sci. USA* 93 6775-9

[18]    Fisher M E and Kolomeisky A B 2001 Simple mechanochemistry describes the dynamics of kinesin molecules *Proc. Natl. Acad. Sci. USA* 98 7748-53

[19]    Astumian R D and Derenyi I 1999 A chemically reversible Brownian motor: application to kinesin

and Ncd *Biophys. J.* 77 993-1002

[20]     Kanada R and Sasaki K 2003 Theoretical model for motility and processivity of two-headed molecular motors *Phys. Rev.* E 67 061917 (1-13)

[21]     Shao Q and Gao Y Q 2006 On the hand-over-hand mechanism of kinesin *Proc. Natl. Acad. Sci. USA* 103 8072-7

[22]     Rice S, Lin A W, Safer D, Hart C L, Naber N, Carragher B O, Cain S M, Pechatnikova E, Wilson-Kubalek E M, Whittaker M, Pate E, Cooke R, Taylor E W, Milligan R A and Vale R D 1999 A structural change in the kinesin motor protein that drives motility *Nature* 402 778-84

[23]     Taniguchi Y, Nishiyama M, Ishii Y and Yanagida T 2005 Entropy rectifies the Brownian steps of kinesin *Nature Chem. Biol.* 1 342-7

[24]     Rice S, Y. C, Sindelar C, Naber N, Matuska M, Vale R D and Cooke R 2003 Thermodynamics properties of the kinesin neck-region docking to the catalytic core *Biophys. J.* 84 1844-54

[25]     Wang Z S, Feng M, Zheng W W and Fan D G 2007 Kinesin is an evolutionarily fine-tuned molecular ratchet-and-pawl device of decisively locked directionality *Biophys. J.* 93 3363-72

[26]     Fan D G, Zheng W W, Hou R, Li F and Wang Z S 2008 Modelling motility of the kinesin dimer from molecular properties of individual monomers *Biochemistry (USA)* 47 4733-42

[27]     Wang Z S 2007 Synergic mechanism and fabrication target for bi-pedal nanomotors *Proc. National Academy of Sciences of the United States of America* 104 17921-6

[28]     Okada Y and Hirokawa N 1999 A processive single-headed motor: Kinesin superfamily protein KIF1A *Science* 283 1152-7

[29]     Mori T, Vale R D and Tomishige M 2007 How kinesin waits between steps *Nature* 450 750-4

[30]     Okada Y and Hirokawa N 2000 Mechanism of the single-headed processivity: Diffusional

anchoring between the K-loop of kinesin and the C terminus of tubulin *Proc. Natl. Acad. Sci. USA* 97 640-5

[31]     Makarov D E, Wang Z, Thompson J B and Hansma H 2002 On the interpretation of force extension curves of single protein molecules *J. Chem. Phys* 116 7760-4

[32]     Guydosh N R and Block S M 2006 Backsteps induced by nucleotide analogs suggest the front head of kinesin is gated by strain *Proc. Natl. Acad. Sci. USA* 103 8054-9

[33]     Block S M, Asbury C L, Shaevitz J W and Lang M J 2003 Probing the kinesin reaction cycle with a 2D optical force clamp *Proceedings of the National Academy of Sciences of the United States of America* 100 2351-6

[34]     Szabo A, Schulten K and Schulten Z 1980 First passage time approach to diffusion controlled reactions *J. Chem. Phys* 72 4350-7

[35]     Wang Z S 2004 A bio-inspired, laser operated molecular locomotive *Phys. Rev.* E 70 031903

[36]     Howard J 2001 *Mechanics of motor proteins and the cytoskeleton* (Sunderland, Massachusetts: Sinauer Associates, Inc.)

[37]     Cross R A 2004 The kinetic mechanism of kinesin *Trends Biochem. Sci.* 29 301-9

[38]     Kolomeisky A B, Stukalin E B and Popov A A 2005 Understanding mechanochemical coupling in kinesins using first-passage-time processes *Physical Review E* 71 031902

[39]     Hackney D D 2005 The tethered motor domain of a kinesin-microtubule complex catalyzes reversible synthesis of bound ATP *Proc. National Academy of Sciences of the United States of America* 102 18338-43

**FIGURE CAPTIONS**

**Fig 1.** Effect of opposing forces on stability (i.e. free energy) of kinesin-MT binding configurations. The insets schematically illustrate the dimer-MT configurations. The labels T and D indicate the ATP- and ADP-bound states of a kinesin head, respectively. The bold red line indicates the zippered state of the neck linkers (represented by curves). The left-hand side points to the plus end of MT, namely the direction for kinesin's motion under physiological condition. The energies for configurations V and VI (shown by dashed lines) are separated from those for configurations I, II and IV (solid lines) by energy gaps much larger than the amount of energy released from ATP hydrolysis. The shadowed area indicates measured values of the stall forces (~ 5.2 – 9 pN) from ref.[14].

**Fig 2.** **A**. Schematic illustration of the major mechanochemical cycle for a forward step under low load. The kinesin-MT binding configurations are labeled as for Fig. 1. **B.** Illustration of the pathways leading to backward stepping under substantial opposing load (indicated by label F). Three possible pathways for detachment of the front head from two-head dimer-MT binding configurations are indicated, which are the hydrolysis-initiated detachment, the pure mechanical detachment of the front head from a nucleotide-free state, and the mechanical detachment from an ATP-bound state.

**Fig 3.** Dependence of stall force on ATP concentration. The filled squares represent the measured values from ref.[14]. The open squares are predictions from this model. The solid line is drawn as a visual guide.

**Fig 4.** Dwell time and velocity versus load. **A** and **B.** Effect of load on dwell time for forward steps (A)

and backward steps (B). **C.** Effect of load on the running velocity of a kinesin dimer. The filled circles and filled triangles are the experimental data from ref.[15] for ATP concentration of 10 $\mu$M and 1 mM, respectively. The solid lines are predictions of the present study for the two values of ATP concentration. Panel **C** also shows the velocity data from ref.[14] for ATP concentration of 2 mM (empty circles). The data can be approximated to be those for 1 mM within an error of less than 15% (see text).

**Fig 5.** Typical trajectories and processivity under backward loads predicted by the present study. **A.** Typical trajectories of both heads of a kinesin dimer (heavy and light lines) under an opposing force of 15 pN at a low ATP concentration of 10 $\mu$M. **B.** The average number of consecutive steps, regardless of direction, that has been effectively made by the kinesin dimer for pre-stall loads and super-stall loads. As the load is further raised beyond ~ 18 pN, any walking will become virtually impossible, and the discussion of processivity is meaningless (see text).

**Fig 6.** Occurring probability for different pathways for the front head detachment from the double-head dimer-MT binding configurations in the super-stall regime.

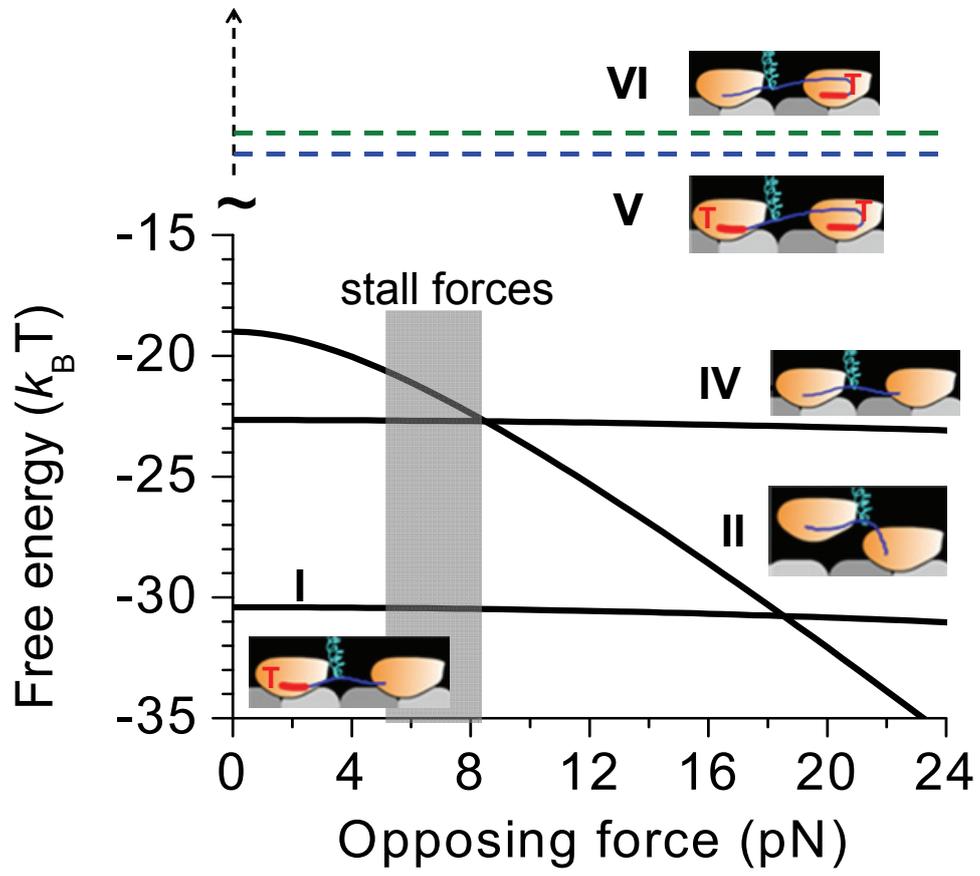

Figure 2, Zheng et al.

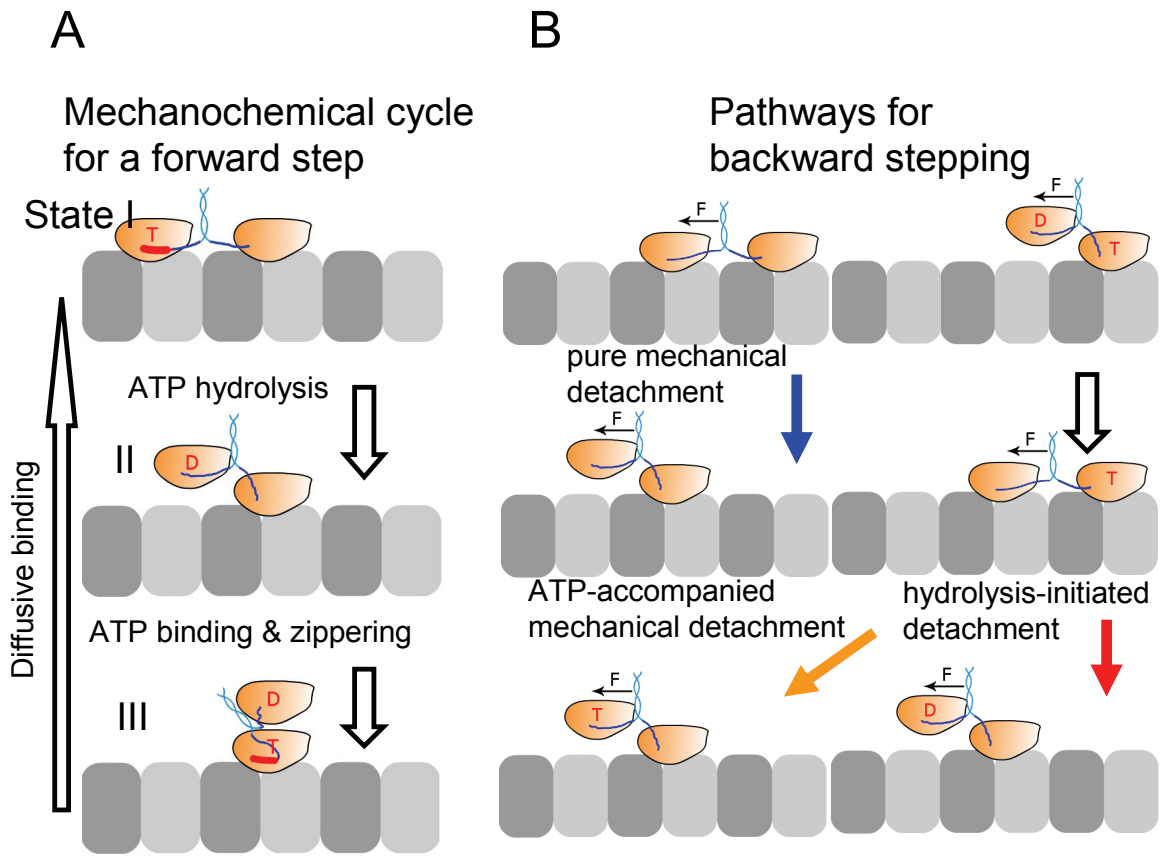

Figure 3, Zheng et al.

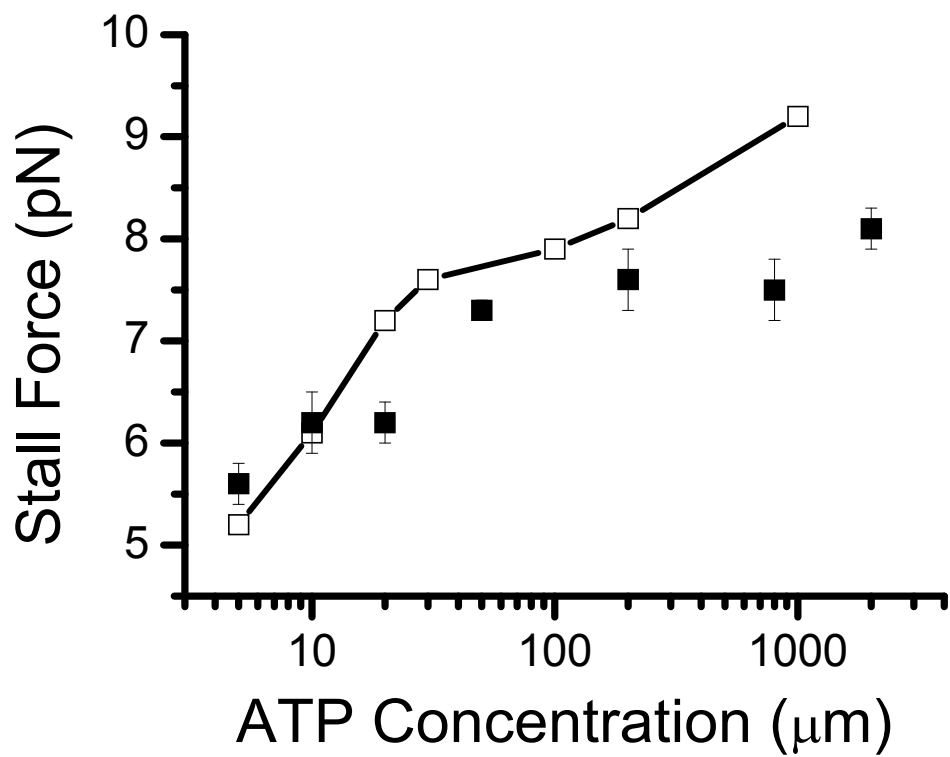

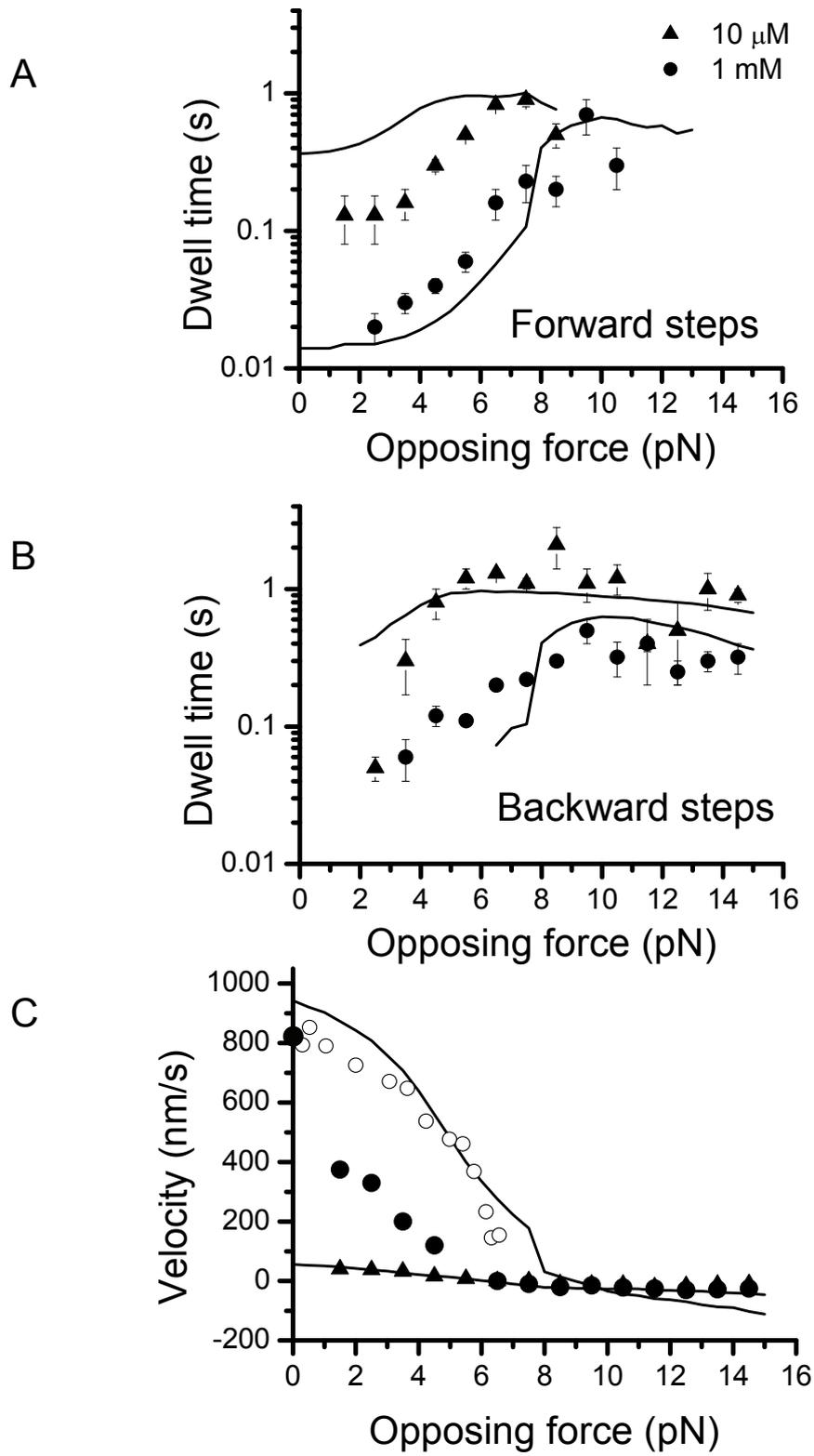

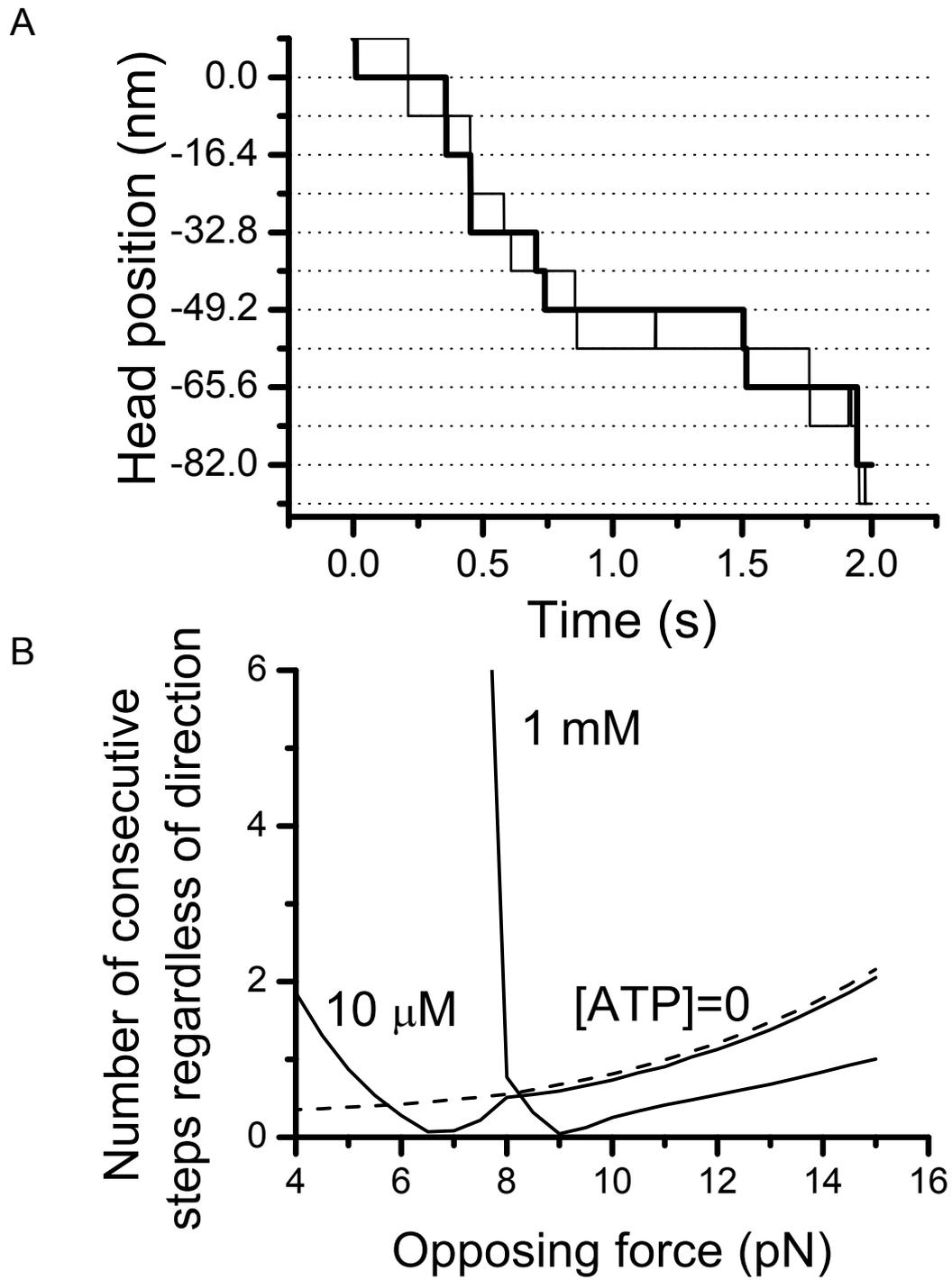

A
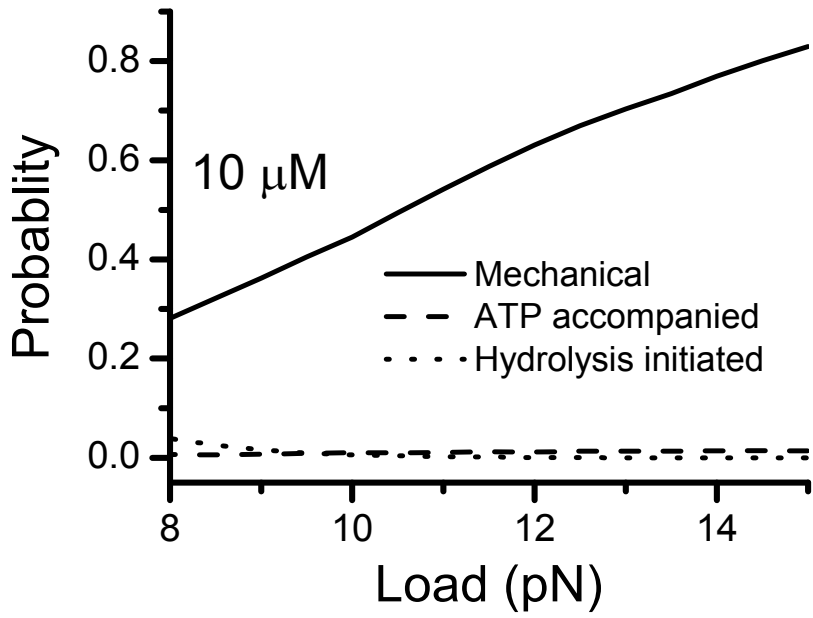

B
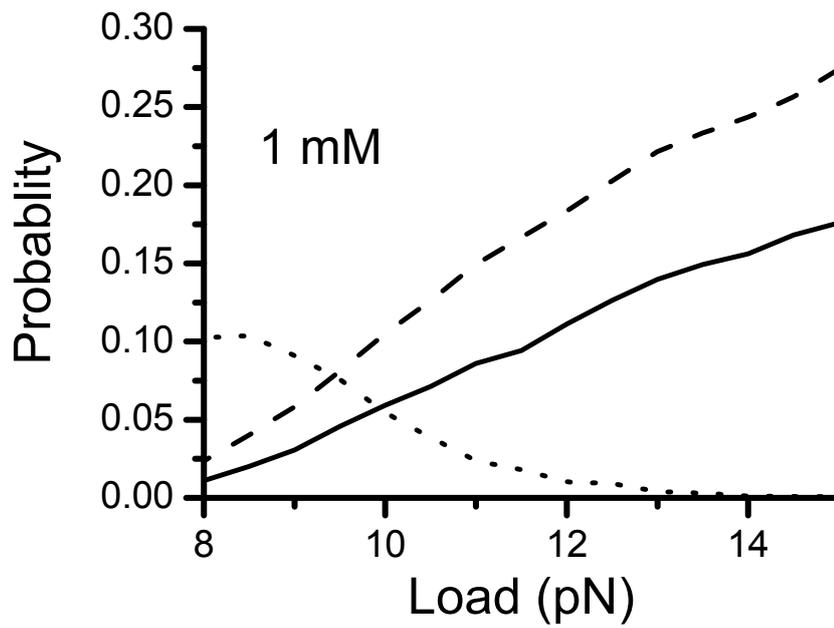